# Bayesian Modeling and Computation for Analyte Quantification in Complex Mixtures Using Raman Spectroscopy

Ningren Han[a,*], Rajeev J. Ram[a]

[a]*Massachusetts Institute of Technology,
77 Massachusetts Avenue, Cambridge, MA 02139, USA*

**Abstract**

In this work, we propose a two-stage algorithm based on Bayesian modeling and computation aiming at quantifying analyte concentrations or quantities in complex mixtures with Raman spectroscopy. A hierarchical Bayesian model is built for spectral signal analysis, and reversible-jump Markov chain Monte Carlo (RJMCMC) computation is carried out for model selection and spectral variable estimation. Processing is done in two stages. In the first stage, the peak representations for a target analyte spectrum are learned. In the second, the peak variables learned from the first stage are used to estimate the concentration or quantity of the target analyte in a mixture. Numerical experiments validated its quantification performance over a wide range of simulation conditions and established its advantages for analyte quantification tasks under the small training sample size regime over conventional multivariate regression algorithms. We also used our algorithm to analyze experimental spontaneous Raman spectroscopy data collected for glucose concentration estimation in biopharmaceutical process monitoring applications. Our work shows that this algorithm can be a promising complementary tool alongside conventional multivariate regression algorithms in Raman spectroscopy-based mixture quantification studies, especially when collecting a large training dataset with high quality is challenging or resource-intensive.

[*]Corresponding author
  *Email addresses:* `ningren@mit.edu` (Ningren Han), `rajeev@mit.edu` (Rajeev J. Ram)

## 1. Introduction

The ability to directly probe the vibrational and rotational state of molecules in the spectral domain has made Raman spectroscopy one of the most widely used tools for chemical and material identification and quantification in physical and biological science. Unlike its close cousins such as near-infrared absorption spectroscopy or fluorescence spectroscopy, Raman spectroscopy of a chemical usually exhibits distinct sharp spectral peaks in the probing region corresponding to various energy transition levels, which some would refer to as the "Raman fingerprint" of the chemical. This high specificity feature is perhaps one of the main reasons for the popularity of Raman spectroscopy despite of its weak signal strength relative to other optical spectroscopy techniques.

Extracting the concentration or quantity information pertaining to certain chemical of interest from a complex mixture spectrum is one of the central themes of modern chemometrics research. Figure 1 shows example spontaneous Raman spectra of a physical mixture as well as its four composition materials, namely glucose, lactic acid, L-lysine, and sodium pyruvate. Concentration or quantity estimation for any or all the composition materials from a set of mixture spectra can be performed with or without any prior knowledge on the Raman spectra of the composition materials. The most widely used analytical methods include supervised multivariate learning algorithms such as partial least squares regression (PLSR) (Wold et al., 2001), principle component regression (PCR) (Næs and Martens, 1988), artificial neural networks (ANNs) (Marini et al., 2008), support vector regression (SVR) (Brereton and Lloyd, 2010) among others (Miller and Miller, 2010, chap. 8). In these methods, a "calibration" process, which is essentially training to the machine learning and applied statistics community, is first carried out for model construction and model selection. The resulting model is subsequently evaluated with cross-validation, bootstrap, or independent test sets.

Despite the general popularity and success of these supervised and training-based multivariate learning algorithms for spectral data analysis, there are limitations that prevent them from being effective or optimal for certain applications. The dependence on the training process means that sufficient mixture spectral data together with the ground truth measurement need to be collected first, possibly in large volume, before a reliable model is built. The process of training data collection itself could be prohibitively expensive or labor-intensive. For example, when using Raman spectroscopy as an on-line tool for monitoring the nutrient and metabolite concentrations in biopharmaceutical processes, the performance of PLSR improves significantly with more training samples at the expense of running the process multiple times (Whelan et al., 2012). In addition, one might need to rerun the training data collection process if certain aspects of the experiment is later modified, e.g. if the growth medium composition is changed in the biopharmaceutical process monitoring example. It is therefore preferable to have an analytical method that can directly perform analyte quantification from the mixture spectrum without a large training pool to begin with in these situations.

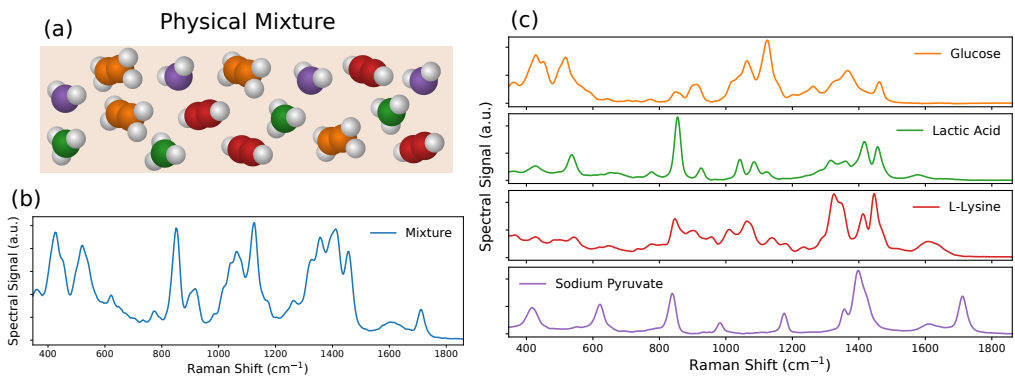

Figure 1: (a) The Raman spectrum contains quantitative information from the mixture molecules under examination. (b) An example Raman spectrum for a physical mixture containing four compositions in (c) at a molar concentration ratio of 1 : 0.47 : 0.66 : 0.35. (c) Raman spectra for the four composition materials in the mixture in (b) measured independently at equal concentrations.

Another common complication typically associated with Raman spectrum processing is baseline estimation and correction (Liland et al., 2010). Baseline signal exists in various forms such as autofluorescence signal from the background material and can exhibit complicated non-linear dynamic behavior due to phenomenon such as autofluorescence photo-bleach and re-covery. This is especially true for biological samples, where the dominant signal comes from intrinsic fluorophores such as NADH, flavin and aromatic amino acids (Afseth et al., 2006; Butler et al., 2016). A number of algorithmic techniques exist to facilitate automatic baseline estimation and correction (Ruckstuhl et al., 2001; Lieber and Mahadevan-Jansen, 2003; Zhao et al., 2007; Galloway et al., 2009; de Rooi and Eilers, 2012; He et al., 2014). However, as mentioned in Moores et al. (2016), most of these methods aim at estimating the baseline signal alone without jointly estimating the peak signals, which may bring potential risks of introducing bias and errors from these separate steps to the estimation.

In this work, we aim at developing an alternative technique to conventional multivariate regression algorithms for analyte quantification in complex mixtures with a Bayesian modeling and computation approach. More specifically, given *a priori* the Raman spectrum measurement of an analyte of interest, which we term as the target analyte in our text, our goal is to quantify its concentration or quantity in a complex mixture spectrum without the need of acquiring additional mixture training data, a scenario that frequently arises in various applications. In addition, the Bayesian approach allows us to simultaneously estimate both the peak and baseline signals, which could mitigate the potential bias and errors with separate estimation steps. There exist several publications aiming at bringing the Bayesian modeling framework to spectral data analysis. Razul et al. (2003); Fischer and Dose (2005); Wang et al. (2008); Nagata et al. (2012); Tokuda et al. (2016) used Bayesian modeling combined with computational methods such as reversible-jump Markov chain Monte Carlo (RJMCMC) or the exchange Monte Carlo method for accurate spectrum variable estimation in various areas such as nuclear emission spectroscopy and mass spectrometry. For Raman spectral data analysis, Zhong et al. (2011) used the Bayesian framework and combined Gibbs and RJMCMC sampler to infer mixture information from a set of multiplexed surface-enhanced Raman spectroscopy (SERS) measurements. Moores et al. (2016) used a sequential Monte Carlo (SMC) sampler for optimal baseline correction and low-concentration analyte quantification. While building upon the common Bayesian modeling and computation principles, our work differs from these prior work due to the fact that our algorithm employs a two-stage processing for quantifying target analyte concentrations in complex mixtures as an alternative to multivariate regression methods such as PLSR with no requirement on pre-existing mixture training data. In Section 2, we provide the statistical framework and computation procedure for our two-stage algorithm, where the first stage is used to learn the peak information for the pure target analyte spectrum and the second stage is for quantifying its concentrations in mixtures. In Section 3, we first demonstrate the utility of this algorithm by testing its performance on a wide range of numerically generated datasets and compare its results with several multivariate regression algorithms. We then report its estimation results on experimental spontaneous Raman spectroscopy data collected for monitoring glucose concentration in biopharmaceutical process with Chinese hamster ovary (CHO) cells, which are the most widely used expression systems for industry production of recombinant protein therapeutics such as monoclonal antibodies used in cancer therapy.

## 2. Methods

### 2.1. Functional Model

Raman spectra are typically collected as one-dimensional signals from a CCD or CMOS detector placed after a dispersive element such as a diffraction grating. Assuming that there are $N$ spectral data points, we model the discrete Raman signal as

$$\mathbf{y} = f_P(\boldsymbol{\nu}) + f_B(\boldsymbol{\nu}) + \boldsymbol{\epsilon}, \tag{1}$$

where $\mathbf{y} \in \mathbb{R}^N$ represents the spectrum array, $\boldsymbol{\nu} \in \mathbb{R}^N$ represents the corresponding Raman shift in wavenumbers, $f_P(\boldsymbol{\nu})$ and $f_B(\boldsymbol{\nu})$ are the functional arrays describing the shape for the Raman peaks and baseline of the signal, and $\boldsymbol{\epsilon} \in \mathbb{R}^N$ is the noise term. $f_P(\boldsymbol{\nu})$ is modeled as the sum of individual Raman peaks each corresponding to an energy transition level as

$$f_P(\boldsymbol{\nu}) = \sum_{j=1}^{k_P} \beta_{P,j} g(\boldsymbol{\nu}; \theta_{P,j}),$$

where $g(\boldsymbol{\nu}; \theta_{P,j})$ is the functional form for the shape of the $j$-th peak with $\theta_{P,j}$ containing the shape variables, and $\beta_{P,j}$ is the corresponding amplitude variable. Depending on the relative contributions from the amplitude correlation time and the coherence lifetime to the effective lifetime of the excited energy states, the functional line shape of a Raman peak can be of the Gaussian profile, the Lorentzian profile, or a combination of both, in which case it can be represented by the Voigt profile (Bradley, 2015). As a popular approximation to the computationally-expensive Voigt profile, the pseudo-Voigt profile uses a linear combination of the Gaussian profile and the Lorentzian profile controlled by a weight factor to adjust their relative contributions (Wertheim et al., 1974). This is what we choose to model the line shape of the Raman peaks in our study. With $l_j$ as the centroid location, $w_j$ as the full width at half maximum (FWHM), and $\rho_j$ as the weight factor for the $j$-th peak, we denote the peak variables for the $j$-th peak as $\theta_{P,j} = (l_j, w_j, \rho_j)$. This leads to

$$g(\boldsymbol{\nu}; \theta_{P,j}) = \rho_j \exp\left\{-\frac{4\ln 2\,(\boldsymbol{\nu} - l_j)^2}{w_j^2}\right\} + (1 - \rho_j)\frac{1}{1 + \left[\frac{2(\boldsymbol{\nu}-l_j)}{w_j}\right]^2}. \quad (2)$$

Meanwhile, the baseline signal $f_B(\boldsymbol{\nu})$ is modeled with a B-spline function, which can be represented as

$$f_B(\boldsymbol{\nu}) = \sum_{j=1}^{k_B} \beta_{B,j} B_{d,j;t}(\boldsymbol{\nu}).$$

Here, $B_{d,j;t}(\boldsymbol{\nu})$ is the $j$-th basis function with degree $d$ and knots $\mathbf{t}$, and can be derived from the Cox-de Boor recursive formula (De Boor et al., 1978). $k_B$ is the number of spline basis functions and $\beta_{B,j}$ is the amplitude coefficient for the $j$-th basis. In our modeling, the knots $\mathbf{t} \in \mathbb{R}^{k_t}$ are chosen as equally-spaced locations in the wavenumber domain and the number of knots $k_t$ satisfies the constraint that $k_t = k_B + d + 1$. In addition, we choose to have a fixed number of spline basis with $k_B = 4$ and set the degree $d$ of the basis function as 3. For Raman spectroscopy, the noise $\boldsymbol{\epsilon}$ may come from a variety of sources including signal shot noise, detector dark current shot noise, temperature and environment fluctuations, laser instability, and so on. With the contributions from these independent sources, we approximate the noise in the observed signal as independent and identically distributed (i.i.d.) Gaussian random noise across the spectral domain.

With the above formulation, Equation 1 can be expressed in a typical Bayesian linear regression form as

$$\mathbf{y} = \mathbf{X}_k(\boldsymbol{\theta}_P)\boldsymbol{\beta}_k + \boldsymbol{\epsilon}, \quad (3)$$

with $\mathbf{y} \in \mathbb{R}^N$, $\boldsymbol{\epsilon} \in \mathbb{R}^N$, $\boldsymbol{\beta}_k = (\boldsymbol{\beta}_P, \boldsymbol{\beta}_B) \in \mathbb{R}^k$, $k = k_P + k_B$ as the overall

model dimension, and $\mathbf{X}_k(\boldsymbol{\theta}_P) \in \mathbb{R}^{N \times k}$ as

$$\mathbf{X}_k(\boldsymbol{\theta}_P) = \begin{bmatrix} g(\nu_1;\theta_{P,1}) & \ldots & g(\nu_1;\theta_{P,k_P}) & B_{d,1;t}(\nu_1) & \ldots & B_{d,k_B;t}(\nu_1) \\ g(\nu_2;\theta_{P,1}) & \ldots & g(\nu_2;\theta_{P,k_P}) & B_{d,1;t}(\nu_2) & \ldots & B_{d,k_B;t}(\nu_2) \\ \vdots & \ddots & \vdots & \vdots & \ddots & \vdots \\ g(\nu_N;\theta_{P,1}) & \ldots & g(\nu_N;\theta_{P,k_P}) & B_{d,1;t}(\nu_N) & \ldots & B_{d,k_B;t}(\nu_N) \end{bmatrix}.$$

With the Gaussian random noise assumption mentioned above, we have

$$\boldsymbol{\epsilon} \sim \mathcal{N}(0, \sigma^2 \mathbf{I}_N),$$

where $\sigma^2$ is the noise variance and $\mathbf{I}_N$ is the identity matrix with dimension $N$.

Given an observed Raman spectrum $\mathbf{y}$, we can jointly estimate the signal decomposition matrix $\mathbf{X}_k(\boldsymbol{\theta}_P)$ as well as the corresponding regression coefficients $\boldsymbol{\beta}_k$ in Equation 3. As the number of Raman peaks $k_P$ is in general not known ahead of the time, model selection is required. We solve this estimation problem by incorporating a hierarchical Bayesian model and using trans-dimensional MCMC computation for model selection and variable estimation.

2.2. Priors

We start solving our model by incorporating Zellner's g-prior (Zellner, 1986), which is a popular choice in Bayesian linear regression and variable selection due to its computational efficiency and the convenience of forming the prior covariance structure from the design matrix itself, into our formulation. The prior for $\boldsymbol{\beta}_k$ is

$$\boldsymbol{\beta}_k | \mathbf{X}_k(\boldsymbol{\theta}_P), g, \sigma^2 \sim \mathcal{N}\left(\boldsymbol{\beta}_{k,0}, g\sigma^2 \left(\mathbf{X}_k^{\intercal}(\boldsymbol{\theta}_P)\mathbf{X}_k(\boldsymbol{\theta}_P)\right)^{-1}\right), \qquad (4)$$

with prior mean $\boldsymbol{\beta}_{k,0}$ and a positive scale variable $g$. Meanwhile, we impose an improper Jeffery's prior on $\sigma^2$ as

$$p(\sigma^2) \propto \sigma^{-2}.$$

Various strategies exist for the modeling of $g$ such as empirical Bayes and fully Bayesian (Liang et al., 2008), here we put an uninformative diffuse inverse-gamma$(\epsilon, \epsilon)$ prior to $g$ as

$$g \sim \mathcal{IG}(a_g, b_g),$$

with $a_g, b_g \to 0$ similar to Razul et al. (2003). This allows a convenient Gibbs update for $g$ due to its conditional conjugacy property.

The number of Raman peaks $k_P$ present in the spectrum is modeled with a Poisson distribution with rate or mean variable $\Lambda$ as [1]

$$k_P | \Lambda \sim \text{Poisson}(\Lambda),$$

and we further model $\Lambda$ with a weak and uninformative conjugate gamma$(\epsilon, \epsilon)$ prior as

$$\Lambda \sim \mathcal{G}a(a_\Lambda, b_\Lambda),$$

with $a_\Lambda, b_\Lambda \to 0$.

Given $k_P$ Raman peaks, we assume conditional independence for the prior distributions for the peak variables in $\boldsymbol{\theta}_P$. With the wavenumber region spanning across $[l_{\min}, l_{\max}]$ and $\Delta l = l_{\max} - l_{\min}$, we assign a uniform flat prior to the locations $\mathbf{l} \in [l_{\min}, l_{\max}]^{k_P}$ of the peaks, which leads to

$$\mathbf{l}|k_P \sim \prod_{i=1}^{k_P} \mathcal{U}(l_i; l_{\min}, l_{\max}) = \left(\frac{1}{\Delta l}\right)^{k_P} \prod_{i=1}^{k_P} \mathbb{1}_{[l_{\min}, l_{\max}]}(l_i).$$

For the widths of the peaks $\mathbf{w} \in \mathbb{R}^{k_P}$, it is desirable to obtain prior information in order to design a suitable prior distribution. As will be described in more details in Section 3.1, we surveyed around 100 Raman peak widths found in common materials and fitted these samples with an inverse-gamma distribution. To account for the limited sample space that we have surveyed and to adopt a conservative approach (Gelman et al., 2008), we intentionally weaken this prior knowledge by scaling the variance of the inverse-gamma distribution by a factor of 4 while keeping the mode of the distribution fixed. With $a_w$ and $b_w$ denoting the parameters corresponding to the scaled inverse-gamma distribution, we have the prior distribution for $\mathbf{w}$ as

$$\mathbf{w}|k_P \sim \prod_{i=1}^{k_P} \mathcal{IG}(w_i; a_w, b_w) = \left(\frac{b_w^{a_w}}{\Gamma(a_w)}\right)^{k_P} \left(\prod_{i=1}^{k_P} w_i^{-a_w-1}\right) \exp\left\{-b_w \sum_{i=1}^{k_P} \frac{1}{w_i}\right\}.$$

---

[1] Although it is more precise to model it as a truncated Poisson distribution due to the finite number of components allowed in our computation, the choice of an untruncated Poisson distribution results in a cleaner conditional posterior distribution without losing much accuracy.

As noted in Bradley (2015), the line shape of the Raman peak can depend on the state of matter of the material due to the impact of the environment on the effective lifetime of the excited energy states for the molecules. For example, solids tend to have Gaussian profiles, gases tend to have Lorentzian profiles, and liquids tend to have features of both. It is therefore possible to assign specific priors for the relative weights $\boldsymbol{\rho} \in [0,1]^{k_P}$ between the Gaussian and Lorentzian profile for the Raman peaks based on knowledge of the material. Here for general purpose, we assign another uninformative flat prior in the range of $[\rho_{\min}, \rho_{\max}]$ with $\rho_{\min} = 0$, $\rho_{\max} = 1$, and $\Delta\rho = \rho_{\max} - \rho_{\min}$ for $\boldsymbol{\rho}$, which leads to

$$\boldsymbol{\rho}|k_P \sim \prod_{i=1}^{k_P} \mathcal{U}(\rho_i; \rho_{\min}, \rho_{\max}) = \left(\frac{1}{\Delta\rho}\right)^{k_P} \prod_{i=1}^{k_P} \mathbb{1}_{[\rho_{\min}, \rho_{\max}]}(\rho_i).$$

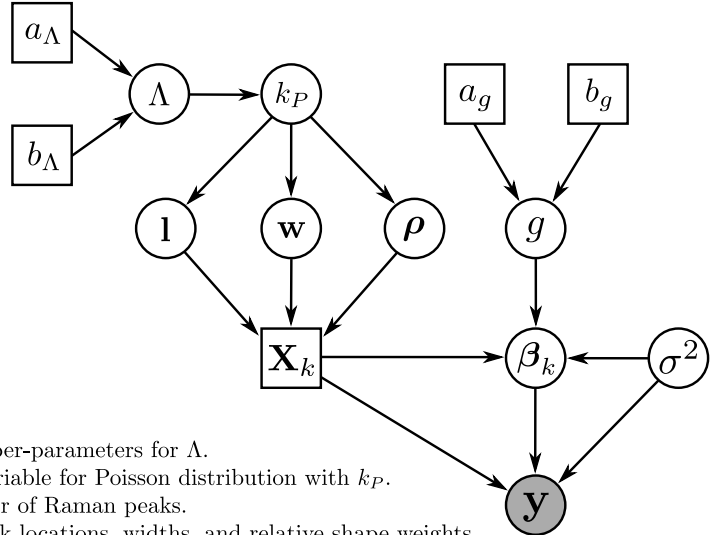

$a_\Lambda, b_\Lambda$: Hyper-parameters for $\Lambda$.
$\Lambda$: Rate variable for Poisson distribution with $k_P$.
$k_P$: Number of Raman peaks.
$\mathbf{l}, \mathbf{w}, \boldsymbol{\rho}$: Peak locations, widths, and relative shape weights.
$\mathbf{X}_k$: Signal decomposition matrix.
$a_g, b_g$: Hyper-parameters for $g$.
$g$: Scale variable for the g-prior.
$\boldsymbol{\beta}_k$: Amplitude variables for the signal decomposition matrix $\mathbf{X}_k$.
$\sigma^2$: Noise variance.
$\mathbf{y}$: Observed spectral array.

Figure 2: Graphical model for the hierarchical Bayesian structure of the spectral signal.

The graphical model representing the hierarchical Bayesian structure of the spectral signal is shown in Figure 2. With the likelihood function of our

model as

$$p\left(\mathbf{y}|\mathbf{X}_k(\boldsymbol{\theta}_P),\boldsymbol{\beta}_k,\sigma^2\right) =$$
$$\frac{1}{\sqrt{|2\pi\sigma^2\mathbf{I}_N|}}\exp\left\{-\frac{1}{2\sigma^2}\left(\mathbf{y}-\mathbf{X}_k(\boldsymbol{\theta}_P)\boldsymbol{\beta}_k\right)^\intercal\left(\mathbf{y}-\mathbf{X}_k(\boldsymbol{\theta}_P)\boldsymbol{\beta}_k\right)\right\},$$

we can express the joint posterior distribution for all the variables as

$$p(g,\sigma^2,\Lambda,k_P,\boldsymbol{\theta}_P,\boldsymbol{\beta}_k|\mathbf{y}) \propto p\left(\mathbf{y}|\mathbf{X}_k(\boldsymbol{\theta}_P),\boldsymbol{\beta}_k,\sigma^2\right)$$
$$p(g)p(\sigma^2)p(\Lambda)p(k_P|\Lambda)p(\mathbf{l}|k_P)p(\mathbf{w}|k_P)p(\boldsymbol{\rho}|k_P)p\left(\boldsymbol{\beta}_k|\mathbf{X}_k(\boldsymbol{\theta}_P),g,\sigma^2\right). \quad (5)$$

2.3. *Bayesian Computation*

No closed-form solution exists for evaluating the joint posterior distribution from Equation 5, we resort to numerical approximation with statistical sampling. In addition, the number of Raman peaks $k_P$ is generally not known beforehand and affects the dimensionality of the variable space $\mathcal{X}_{k_P}$ for the model. Therefore, model selection across the model space with different $k_P$ is required. A diversity of criteria and methodologies exists for Bayesian model selection (Wasserman, 2000). Here we adopt a unified approach for joint variable estimation and model selection with the RJMCMC technique introduced by Green (1995); Richardson and Green (1997). The RJMCMC method samples from the union space $\mathcal{X} = \cup_{k_P\in\mathcal{K}}(\{k_P\}\times\mathcal{X}_{k_P})$ for the potential models, where $\mathcal{K}$ in our case is a countable set containing all the possible Raman peak number in the spectrum, by constructing a reversible Markov chain in the general state space. The trans-dimensional moves across the models in RJMCMC can be incorporated inside the general Metropolis-Hastings paradigm in a straightforward manner. With marginalization, the posterior probability of being in any variable space $\mathcal{X}_{k_P}$ can be obtained, and model selection can be performed accordingly. For more in-depth discussions on the model determination aspects with RJMCMC, Hastie and Green (2012) serves as an excellent reference.

In the following text, we use $|\ldots$ to denote conditioning on all other random variables. With the hierarchical Bayesian structure imposed by our model, several variables can be conveniently updated with Gibbs sampling. They are $g$, $\Lambda$, $\sigma$, and $\boldsymbol{\beta}_k$. The corresponding Gibbs updates are shown in Appendix A. The conditional posterior distribution for the rest of the variables $\boldsymbol{\theta}_P$ and $k_P$ does not admit a tractable form. To sample $\boldsymbol{\theta}_P$ and $k_P$, we first integrate out $\sigma^2$ and $\boldsymbol{\beta}_k$ from the full joint posterior distribution for simplification. This leaves the conditional posterior probability for $\boldsymbol{\theta}_P$ and $k_P$ known to a proportionality as

$$p(\boldsymbol{\theta}_P, k_P | \ldots) \propto \frac{\Lambda^{k_P}}{k_P!} \tilde{b}_{\sigma^2}^{-\frac{N}{2}} \frac{1}{\sqrt{|(g+1)\mathbf{I}_k|}} \left(\frac{1}{\Delta l}\right)^{k_P} \prod_{i=1}^{k_P} \mathbb{1}_{[l_{\min}, l_{\max}]}(l_i)$$

$$\left(\frac{b_w^{a_w}}{\Gamma(a_w)}\right)^{k_P} \left(\prod_{i=1}^{k_P} w_i^{-a_w-1}\right) \exp\left\{-b_w \sum_{i=1}^{k_P} \frac{1}{w_i}\right\} \left(\frac{1}{\Delta \rho}\right)^{k_P} \prod_{i=1}^{k_P} \mathbb{1}_{[\rho_{\min}, \rho_{\max}]}(\rho_i). \tag{6}$$

where $\tilde{b}_{\sigma^2}$ is defined in Equation A.1. Samples of $\boldsymbol{\theta}_P$ and $k_P$ can be obtained with the RJMCMC method, which can be viewed as a generalization of the Metropolis-Hastings algorithm with additional trans-dimensional moves. Denoting the current variable state as $\mathbf{x} = (\boldsymbol{\theta}_P, k_P)$, for any proposed variable state $\mathbf{x}' = (\boldsymbol{\theta}'_P, k'_P)$, we can calculate the corresponding Metropolis-Hastings acceptance probability $a(\mathbf{x}, \mathbf{x}') = \min\{1, A(\mathbf{x}, \mathbf{x}')\}$, where $A(\mathbf{x}, \mathbf{x}')$ can be calculated for each move type respectively.

Under the RJMCMC sampling scheme, in addition to the regular Metropolis-Hastings within-dimensional moves, trans-dimensional reversible move pairs also need to be devised. With proper engineering, by generating assistive random variable $\mathbf{u}$ from proposal density $g(\mathbf{u})$, the proposed state $\mathbf{x}'$ can be constructed by a deterministic function $h(\cdot)$ as $(\mathbf{x}', \mathbf{u}') = h(\mathbf{x}, \mathbf{u})$, where $\mathbf{u}'$ is a random variable that can be generated from proposal density $g'(\mathbf{u}')$ so that one can reversely jump from $\mathbf{x}'$ back to $\mathbf{x}$ based on the inverse of $h(\cdot)$. The transformation $h(\cdot)$ needs to be a diffeomorphism with matching dimensions for $(\mathbf{x}, \mathbf{u})$ and $(\mathbf{x}', \mathbf{u}')$, which means $n_\mathbf{x} + n_\mathbf{u} = n_{\mathbf{x}'} + n_{\mathbf{u}'}$ with $n$ being the variable dimension. Let $m$ and $m'$ be the indices for reversible move pairs across the dimensions of $\mathbf{x}$ and $\mathbf{x}'$ in set $\mathcal{M}$ containing all the possible moves and $q(m|\mathbf{x})$ be the probability of taking move $m$ at state $\mathbf{x}$, we can calculate $A(\mathbf{x}, \mathbf{x}')$ as

$$A(\mathbf{x}, \mathbf{x}') = \frac{p(\mathbf{x}')q(m'|\mathbf{x}')g'(\mathbf{u}')}{p(\mathbf{x})q(m|\mathbf{x})g(\mathbf{u})} \left|\frac{\partial(\mathbf{x}', \mathbf{u}')}{\partial(\mathbf{x}, \mathbf{u})}\right|,$$

where $|\cdot|$ denotes the determinant of the transformation Jacobian.

In this work, we employ four trans-dimensional moves to facilitate cross model mixing, where similar strategies can be found in applications such as Bayesian mixture estimation (Richardson and Green, 1997). These moves are:

1. Birth of a new peak.
2. Death of an existing peak.
3. Split of an existing peak.
4. Merge of two adjacent peaks.

The RJMCMC move acceptance probabilities for the within-dimensional move as well as each of these moves are shown in Appendix B.

With the hybrid Gibbs and RJMCMC sampling schedules described above, we can describe an algorithm for joint peak variable and baseline estimation with a Raman spectrum. At each sampling iteration, the RJMCMC move for this iteration is first determined with a categorical random variable $m$ with the support corresponding to the indices of the available moves in $\mathcal{M}$. $\boldsymbol{\theta}_P$ and $k_P$ are updated subsequently based on the move type $m$. This essentially creates a combined mixture MCMC transition kernel for the update of $\boldsymbol{\theta}_P$ and $k_P$. Afterwards, $(g, \Lambda, \sigma^2, \boldsymbol{\beta}_k)$ are updated with Gibbs sampling. Once the Markov chain is fully mixed, model selection based on $k_P$ can be performed. For example, the *maximum a posteriori* (MAP) estimation for $k_P$ can be obtained as

$$\hat{k}_P = \arg\max_{k_P} p(k_P|\mathbf{y}).$$

For spectra having visually distinct and well-spaced peaks, the above Bayesian sampling schedule works well with a fixed and equally-likely move proposal distribution for $m$. However, for more complex spectra having a large number of peaks that may have tightly spaced or partially overlapping peaks, we notice that frequent label switching caused by trans-dimensional moves in steady state can become problematic for variable estimation (Jasra et al., 2005). In addition, in these situations, during the early inter-state mixing iterations, negative amplitudes can be assigned to some peaks (while still maintaining an overall spectral signal match with the observed spectrum). These peaks with negative amplitudes may stay throughout the iterations, which create unphysical decomposition results. We address these two problems with the following approaches.

As a solution to the first problem, we employ a heuristic approach by gradually decreasing the probability of taking trans-dimensional moves throughout the iterations. At iteration $i$, the move type is determined from $m^{(i)}$ sampling from the categorical proposal distribution $p_m^{(i)}(m)$ with probability mass as $(p_w^{(i)}, p_b^{(i)}, p_d^{(i)}, p_s^{(i)}, p_m^{(i)})$ for each move – $p_w^{(i)}$ corresponds to the within move, $p_b^{(i)}$ and $p_d^{(i)}$ correspond to the birth and death moves, and $p_s^{(i)}$ and $p_m^{(i)}$ correspond to the split and merge moves. Using $p_b^{(i)}$ as an example, we adjust its value in each iteration as

$$p_b^{(i)} = \left(p_b^{(0)}\right)^{1/T^{(i)}},$$

with $p_w^{(0)} = p_b^{(0)} = p_d^{(0)} = p_s^{(0)} = p_m^{(0)}$, $T^{(0)} = 1$, $\lim_{i \to \infty} T^{(i)} = 0$, and a linearly decreasing cooling schedule for $T^{(i)}$. We perform this adjustment for all the trans-dimensional moves. Meanwhile, we increase the within-dimensional move probability accordingly with the constraint that $p_w^{(i)} + p_b^{(i)} + p_d^{(i)} + p_s^{(i)} + p_m^{(i)} = 1$. This treatment is similar to simulated annealing, and effectively creates a non-homogeneous Markov chain in the general state space (Andrieu et al., 2003). Convergence results can be obtained with simulated annealing-like algorithms (Van Laarhoven and Aarts, 1987), which we do not pursue in this work. Once the steady state is reached, all samples are effectively drawn from the same model with $\hat{k}_P$. Therefore, variable values can be estimated without explicitly performing model selection.

For the second problem, we enforce a non-negativity constraint on the amplitude coefficients $\boldsymbol{\beta}_P$ for the peaks. During the sampling iterations, if any peak with a negative amplitude is generated, we discard the sample and restart the current iteration step until all peaks are of non-negative values. This emulates rejection sampling and effectively adds a non-negative support constraint on $\boldsymbol{\beta}_P$ for the prior and posterior probability distributions in Equation 4 and Equation A.2. We note here that even without the annealing schedule described above, this re-sampling operation is only mostly required during early iterations where the computation is rapidly converging in the model domain. Once the model dimension is reasonably converged, negative peak amplitude generation seldomly occurs.

Denoting $\boldsymbol{\phi}_{\hat{k}_P} = (g, \Lambda, \sigma^2, \boldsymbol{\beta}_k, \boldsymbol{\theta}_P)$ for the variables associated with $\hat{k}_P$, we can estimate the conditional posterior expected values for $\boldsymbol{\phi}_{\hat{k}_P}$ as

$$\mathbb{E}_{p(\boldsymbol{\phi}_{\hat{k}_P}|\mathbf{y},\hat{k}_P)}[\boldsymbol{\phi}_{\hat{k}_P}] \approx \frac{1}{M} \sum_{i=1}^{M \to \infty} \boldsymbol{\phi}_{\hat{k}_P}^{(i)} \qquad (7)$$

with $\boldsymbol{\phi}_{\hat{k}_P}^{(i)}$ being the $i$-th sample drawn from the conditional posterior distribution $p(\boldsymbol{\phi}_{\hat{k}_P}|\mathbf{y}, \hat{k}_P)$ and $M$ being the total number of samples. Alternative estimation criterion such as the MAP estimator can also be used here.

*2.4. Two-Stage Algorithm*

The above Bayesian formulation and computation provide a framework to simultaneously estimate the peak and baseline signal in a Raman spectrum. In order to further it into a quantification algorithm applicable to practical scenarios, we propose a two-stage algorithm built upon this framework.

In many analyte quantification tasks involving Raman spectroscopy, the goal is to quantify one or more target analyte in mixture spectra. For simplicity of presentation we restrict our attentions to one target analyte but note that the extension to multiple target analytes is straightforward. We also assume an aqueous mixture environment in our analysis. While not required by many multivariate regression techniques such as PLSR, the actual spectrum of the target analyte is often easy to acquire through a separate reference measurement. Our algorithm takes advantage of this aspect by first learning the peak representation for the target analyte at a known concentration. This can be achieved through the Bayesian computation process described above working on the reference target analyte spectrum. Once the peak variables for the target analyte are learned in this first stage, we move on to the second stage with the mixture spectrum where the concentration for the target analyte in the mixture needs to be determined. In this stage, the processing is slightly modified relative to the first stage to take into account of the learned representation of the target analyte. The modifications are described as follows.

With the peak and baseline decomposition for the reference target analyte spectrum in the first stage as in Equation 1, and $\hat{\boldsymbol{\theta}}_P$ and $\hat{\boldsymbol{\beta}}_P$ corresponding to the estimated peak variables for the target analyte according to Equation 7, we define the Raman peak signal at unit concentration for the target analyte as

$$\tilde{f}_P(\boldsymbol{\nu}) = \frac{f_P(\boldsymbol{\nu})}{c_{\text{pure}}} = \sum_{j=1}^{k_P} \frac{\hat{\beta}_{P,j}}{c_{\text{pure}}} g(\boldsymbol{\nu}; \hat{\theta}_{P,j}), \tag{8}$$

where $c_{\text{pure}}$ is the target analyte concentration in the reference measurement.

In the second stage, the observed signal in the mixture spectrum now can be modeled as

$$\mathbf{y} = f_T(\boldsymbol{\nu}) + f_I(\boldsymbol{\nu}) + f_B(\boldsymbol{\nu}) + \boldsymbol{\epsilon},$$

where here $f_T(\boldsymbol{\nu})$ represents peaks originating from the target analyte, $f_I(\boldsymbol{\nu})$ represents peaks from the other analytes in the mixture, which we call the

*First stage*
**Input** : Reference spectrum of the target analyte
Initialize $T^{(0)}$, $(p_w^{(0)}, p_b^{(0)}, p_d^{(0)}, p_s^{(0)}, p_m^{(0)})$, and the spectrum variables
**for** $i \leftarrow 1$ **to** $I$ **do**
> Determine move type $m^{(i)}$ with $(p_w^{(i)}, p_b^{(i)}, p_d^{(i)}, p_s^{(i)}, p_m^{(i)})$
> Based on move type $m^{(i)}$, sample $\boldsymbol{\theta}_P$ with RJMCMC and update $k_P$
> Sample $g$, $\Lambda$, $\sigma^2$, $\boldsymbol{\beta}_k = (\boldsymbol{\beta}_P, \boldsymbol{\beta}_B)$ with Gibbs sampling
> **if** $\exists \beta_{P,j} \in \boldsymbol{\beta}_P$ s.t. $\beta_{P,j} < 0$ **then**
>> Discard current samples and restart current iteration
>
> **end**
> Update $i, T^{(i)}, (p_w^{(i)}, p_b^{(i)}, p_d^{(i)}, p_s^{(i)}, p_m^{(i)})$

**end**
Estimate $\hat{\boldsymbol{\theta}}_P$, $\hat{\boldsymbol{\beta}}_P$, and calculate $\tilde{f}_P(\boldsymbol{\nu})$

*Second stage*
**Input** : Mixture spectrum
Initialize $T^{(0)}$, $(p_w^{(0)}, p_b^{(0)}, p_d^{(0)}, p_s^{(0)}, p_m^{(0)})$, and the spectrum variables
**for** $i \leftarrow 1$ **to** $I$ **do**
> Determine move type $m^{(i)}$ with $(p_w^{(i)}, p_b^{(i)}, p_d^{(i)}, p_s^{(i)}, p_m^{(i)})$
> Based on move type $m^{(i)}$, sample $\boldsymbol{\theta}_I$ with RJMCMC and update $k_I$
> Sample $g$, $\Lambda$, $\sigma^2$, $\boldsymbol{\beta}_k = (c_{\text{mix}}, \boldsymbol{\beta}_I, \boldsymbol{\beta}_B)$ with Gibbs sampling
> **if** $\exists \beta_{I,j} \in \boldsymbol{\beta}_I$ s.t. $\beta_{I,j} < 0$ **then**
>> Discard current samples and restart current iteration
>
> **end**
> Update $i, T^{(i)}, (p_w^{(i)}, p_b^{(i)}, p_d^{(i)}, p_s^{(i)}, p_m^{(i)})$

**end**
Estimate $\hat{c}_{\text{mix}}$

**Algorithm 1:** The two-stage algorithm for analyte quantification in mixture spectrum with Bayesian modeling and computation.

interfering analytes, and the rest follows as previous. The target analyte signal $f_T(\boldsymbol{\nu})$ is related to its concentration in the mixture $c_{\text{mix}}$ as

$$f_T(\boldsymbol{\nu}) = c_{\text{mix}} \tilde{f}_P(\boldsymbol{\nu}).$$

In order to solve for $c_{\text{mix}}$, a similar Bayesian computation process relative to the first stage can be carried out except for that $\tilde{f}_P(\boldsymbol{\nu})$ is kept as a fixed basis with estimated $\hat{\boldsymbol{\theta}}_P$ and $\hat{\boldsymbol{\beta}}_P$ as in Equation 8. With $\boldsymbol{\theta}_I$ as the peak variables for the interfering analytes, this means that $\mathbf{X}_k(\boldsymbol{\theta}_I) \in \mathbb{R}^{N \times k}$, now depends on $\boldsymbol{\theta}_I$, is

$$\mathbf{X}_k(\boldsymbol{\theta}_I) = \begin{bmatrix} \tilde{f}_P(\nu_1) & g(\nu_1; \theta_{I,1}) & \ldots & g(\nu_1; \theta_{I,k_I}) & B_{d,1;t}(\nu_1) & \ldots & B_{d,k_B;t}(\nu_1) \\ \tilde{f}_P(\nu_2) & g(\nu_2; \theta_{I,1}) & \ldots & g(\nu_2; \theta_{I,k_I}) & B_{d,1;t}(\nu_2) & \ldots & B_{d,k_B;t}(\nu_2) \\ \vdots & \vdots & \ddots & \vdots & \vdots & \ddots & \vdots \\ \tilde{f}_P(\nu_N) & g(\nu_N; \theta_{I,1}) & \ldots & g(\nu_N; \theta_{I,k_I}) & B_{d,1;t}(\nu_N) & \ldots & B_{d,k_B;t}(\nu_N) \end{bmatrix}$$

to take into consideration of the target analyte spectrum. Correspondingly, $\boldsymbol{\beta}_k = (c_{\text{mix}}, \boldsymbol{\beta}_I, \boldsymbol{\beta}_B) \in \mathbb{R}^k$, $k_I$ represents the number of peaks coming from the interfering analytes in the mixture, and $k = k_I + k_B + 1$. Afterwards, all the variable sampling schedule and estimation procedure from the previous section can be directly applied to estimate $\hat{c}_{\text{mix}}$ and the rest of the variables. The overall algorithm is shown in Algorithm 1.

## 3. Results

### 3.1. Numerical Experiment Setup

We first set up the numerical experiment environment for exploring the performance of our algorithm under various situations. For all the simulated spectra, the Raman shift wavenumber range spanned across 400 cm$^{-1}$ to 1600 cm$^{-1}$ and contained 300 equally-spaced spectral points. We simulated our studies under the same use case as in actual practice where a reference measurement of the Raman spectrum for the target analyte at a known concentration was first given and our goal was to quantify its concentration in mixture measurements in the presence of other interfering analytes at unknown concentrations.

For any simulated analyte including the target analyte, we modeled its Raman spectrum at unit concentration by explicitly generating the Raman peaks. The number of Raman peaks $k_P$ for the analyte was first determined. Afterwards, we generated the corresponding peak random variables $\boldsymbol{\theta}_P = (\mathbf{l}, \mathbf{w}, \boldsymbol{\rho}, \mathbf{a})$ from predefined probability distributions. Here $\mathbf{l}$, $\mathbf{w}$, and $\boldsymbol{\rho}$ are defined in the previous section and $\mathbf{a} \in [0,1]^{k_P}$ are the peak amplitudes at unit concentration. For all the peaks, $\mathbf{l}$ were independently and uniformly generated across the available spectrum span, and $\boldsymbol{\rho}$ and $\mathbf{a}$ were both independently and uniformly generated from range $[0, 1]$. For $\mathbf{w}$ to be representable to actual Raman peaks encountered in practice, we surveyed around 100 Raman peaks from 11 common materials including phenylalanine, tryptophan, tyrosine, alanine, glycine, glucose, lactic acid, acetic acid, succinic acid, ethanol and water. We extracted the Raman peaks and their widths and fitted these width samples with an inverse-gamma distribution. The histogram for the peak width samples and the fitted probability density function (PDF) are shown in Figure 3 (a). We denote this PDF as $p_g(\mathbf{w})$ and sampled $\mathbf{w}$ independently from this probability distribution in our simulations. Once $k_P$ and $\boldsymbol{\theta}_P$ were both determined, the Raman spectrum at unit concentration could be represented as

$$\tilde{f}_P(\boldsymbol{\nu}) = \sum_{j=1}^{k_P} a_j g(\boldsymbol{\nu}; \theta_{P,j})$$

with $g(\boldsymbol{\nu}; \theta_{P,j})$ defined in Equation 2. Given $\tilde{f}_P(\boldsymbol{\nu})$ for each analyte, we could generate any mixture spectrum by adding together the spectral signals from all the constituent analytes adjusted linearly by their respective concentrations in the mixture. In addition, we also added the baseline signal represented by a low-order polynomial curve, where for each baseline curve, small random perturbations were added to the fitting points used to generate the polynomial curve to ensure a varying baseline across all the spectra. At last, we added independent and additive Gaussian random noise with standard deviation $\sigma$ across the array to generate the final spectra. For the reference target analyte spectrum, no spectral signal from any other analyte was added, but we still included the baseline signal and the noise to resemble an actual measurement.

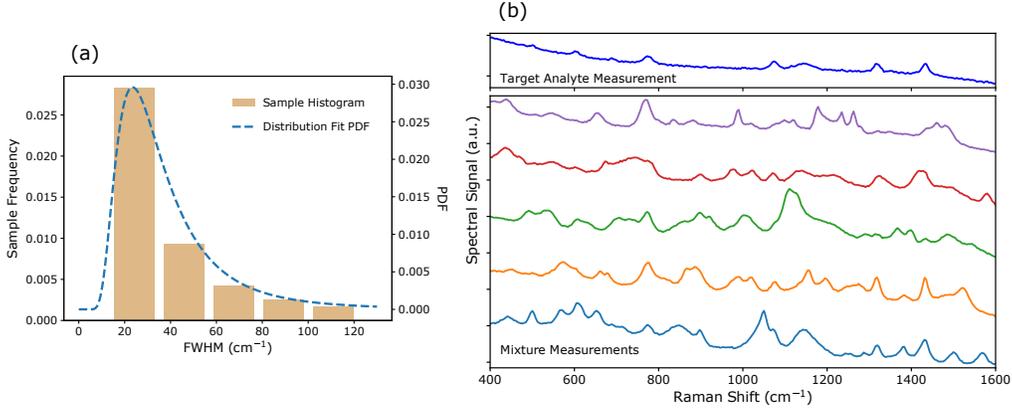

Figure 3: (a) Histogram for the widths of around 100 Raman peaks surveyed in common materials and the PDF plot for an inverse-gamma probability distribution fit. This PDF was used to generate the simulated Raman spectra in our studies. (b) Examples of the simulated target analyte spectrum and 5 mixture spectra each with 5 randomly generated interfering analytes. $\sigma = 1$ for all these simulated measurements.

In the following numerical experiments, we fixed the target analyte spectrum with $k_P = 10$ randomly generated Raman peaks across all the mixture studies. For each mixture, the number of co-existing interfering analytes is denoted as $N_I$. The number of Raman peaks was set as 10 for each interfering analyte similar to that of the target analyte. The concentration for each analyte in the mixture including the target analyte was uniformly and randomly generated from range $[0, 60]$. For the reference target analyte spectrum generation, we set its concentration at 30 and noise scale $\sigma$ at 1. The mixture spectra were all randomly and independently generated from the process described above. Sample plots for the target analyte spectrum as well as 5 mixture spectra each with 5 random interfering analytes are shown in Figure 3 (b). $\sigma = 1$ for all the spectra in the figure.

## 3.2. Mixture Environment Study

With the above settings, we were able to validate our algorithm with simulated target analyte and mixture spectra. As an illustrative example, we show an estimation result in Figure 4. Figure 4 (a) and (b) show the baseline estimation and peak decomposition results for a simulated reference target analyte spectrum. The estimated peak variables obtained in this step were further used to quantify the target analyte concentrations in mixtures, as shown in Figure 4 (c) and (d). With the mixture spectrum, other than the amplitude coefficient of the learned target analyte, a variable number of Raman peaks were also fitted with the RJMCMC computation to explain the peak signals from the rest of the interfering analytes. This ensured that all the peaks appearing in the mixture spectrum were properly assigned to either the target analyte or the interfering analytes, resulting in the most likely amplitude estimation for the target analyte peaks in the mixture spectrum. This in turn corresponded to the concentration of the target analyte in the mixture. In the simulation as shown in Figure 4, $\sigma = 1$ for the target analyte spectrum, $N_I = 5$ and $\sigma = 1$ for the mixture spectrum. The concentration

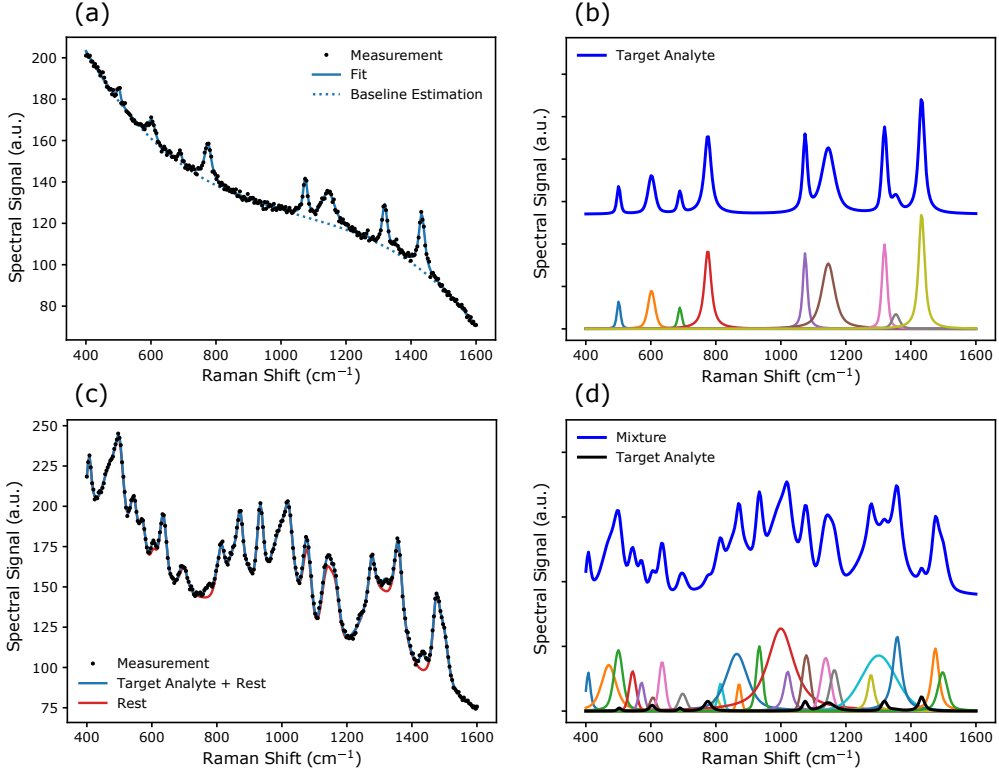

Figure 4: (a) and (b) Example plots for a simulated target analyte spectrum, and its baseline estimation and peak decomposition results with our algorithm. (c) and (d) Example plots for analyte assignment and peak decomposition for a mixture spectrum based on peak variables obtained from (a) and (b) for the target analyte. The resulting spectrum amplitude for the target analyte was subsequently used to estimate its concentration in the mixture. In this simulation, $\sigma = 1$ for the target analyte spectrum, $N_I = 5$ and $\sigma = 1$ for the mixture spectrum. The concentration for the target analyte in the mixture was 5 and the estimated concentration from our algorithm was 4.6

for the target analyte in the mixture was 5 and the estimated concentration from our algorithm was 4.6.

Next we evaluated the generalized performance of our algorithm when the number of co-existing interfering analytes $N_I$ and the additive Gaussian noise scale $\sigma$ varied. The noise scale $\sigma$ was kept as 1 for the reference target analyte spectrum. For each mixture test set with a fixed $N_I$ and $\sigma$, we generated 1000 mixture spectra with randomly generated interfering analytes and randomly generated concentrations for all the constituent analytes as described previously. We used the root mean squared error (RMSE) between our estimations and the ground truth values across all the 1000 measurements as the performance metric. In total, we generated 35 test sets where we varied $N_I$ from 1 to 7 and $\sigma$ from 1 to 5, both in steps of 1. The resulting RMSE across the 1000 test spectra for each set is shown in Table 1 and the corresponding 2D heatmap is shown in Figure 3 (a). In addition, 1D plots showing how RMSE changes with $N_I$ under fixed $\sigma$ (and vice versa) are shown in Figure 3 (b).

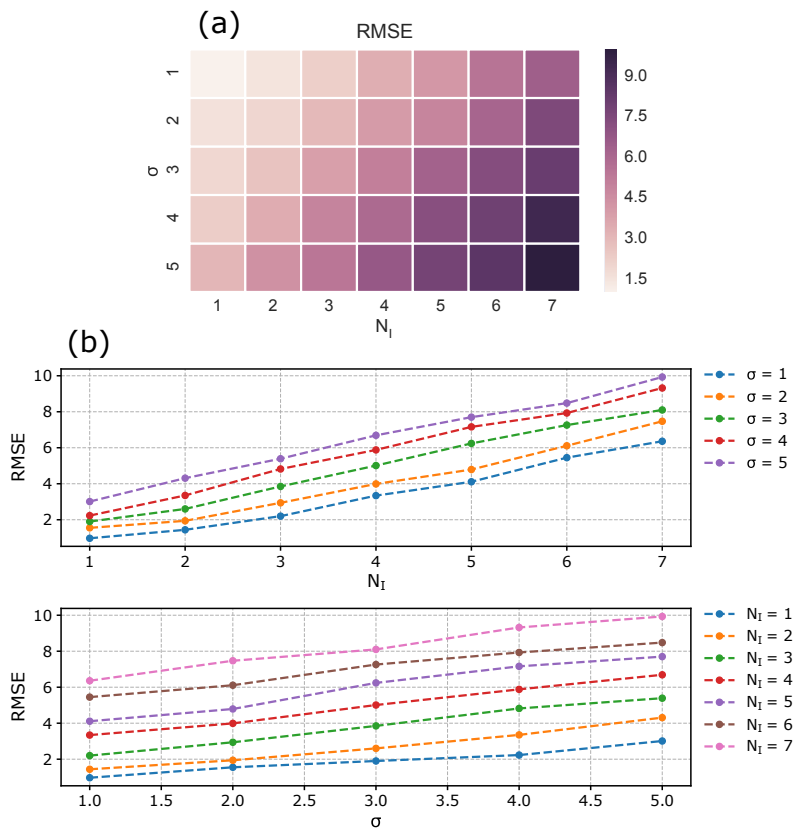

Figure 5: (a) RMSE heatmap plot with varying $N_I$ and $\sigma$. (b) 1D plots showing the RMSE change with fixed $\sigma$ (above) and fixed $N_I$ (below). $N_I$ is the number of co-existing interfering analytes in the mixture and $\sigma$ is the standard deviation of the additive Gaussian random noise. All RMSEs were calculated based on 1000 independently generated random mixtures.

As seen from these results, overall our algorithm was able to provide a consistent and reliable estimation for the target analyte concentration over a wide range of test conditions. The RMSEs under all test cases were below $\approx 17\%$ of the concentration variation from 0 to 60 for the target analyte. As the measurement became noisier or more interfering analytes were added to the mixture, more estimation error was observed. This is expected as the effect of more noise or more interfering analytes increases the uncertainty of correct analyte peak assignment and peak amplitude estimation. A closer examination of error change with fixed $N_I$ or fixed $\sigma$ in Figure 5 (b) indicates a linear increase of error with the other variable. This suggests that a simple linear model with RMSE being the dependent variable, and $N_I$ and $\sigma$ being the independent variables may be able to predict our algorithm's performance in a more generalized scenario. However, further research is needed to rigorously analyze the error bound of our algorithm under these situations.

| $N_I$ | 1 | 2 | 3 | 4 | 5 | 6 | 7 |
|---|---|---|---|---|---|---|---|
| $\sigma = 1$ | 1.0 | 1.4 | 2.2 | 3.3 | 4.1 | 5.5 | 6.4 |
| $\sigma = 2$ | 1.6 | 1.9 | 2.9 | 4.0 | 4.8 | 6.1 | 7.5 |
| $\sigma = 3$ | 1.9 | 2.6 | 3.9 | 5.0 | 6.2 | 7.3 | 8.1 |
| $\sigma = 4$ | 2.2 | 3.4 | 4.8 | 5.9 | 7.2 | 7.9 | 9.3 |
| $\sigma = 5$ | 3.0 | 4.3 | 5.4 | 6.7 | 7.7 | 8.5 | 9.9 |

Table 1: RMSE values with varying $N_I$ and $\sigma$ (also plotted in Figure 5). $N_I$ is the number of co-existing interfering analytes in the mixture and $\sigma$ is the standard deviation of the additive Gaussian random noise. All RMSEs were calculated based on 1000 independently generated random mixtures.

### 3.3. Comparison Study

We further compared the performance of our algorithm against several popular multivariate regression quantification algorithms in chemometrics. Three algorithms including partial least squares regression (PLSR), principle component regression (PCR) and ridge regression (RR) were selected for this comparison study. The implementations for these algorithms were from the scikit-learn 0.19.1 package with Python 3.6. An important distinction between these multivariate regression algorithms and our algorithm is that they are typically built based on a mixture training set with cross validation for model selection. This requires pre-existing mixture spectra as well as the corresponding ground truth reference measurements for the target analyte in the mixtures. In practice, the ground truth reference measurements are usually obtained through a separate chemical assay such as high performance liquid chromatography (HPLC). On the contrary, our algorithm only requires the reference spectrum of the target analyte at a known concentration as prior information before the actual estimation.

Our focus in this study is to investigate how the quantification results compare across the different algorithms as the number of mixture training data varies. This comparison study is relevant for practical applications as mixture training data itself is often labor or resource intensive to acquire in large volume since in addition to spectral data collection, special sample handling or additional analytical measurement like HPLC is usually required. In contrast, the reference spectrum of the target analyte required by our algorithm is much easier to prepare in practice. For the three multivariate regression algorithms, we created training and test sets with a randomized process similar to previously described, except for that now we kept the same fixed mixture components across each training/test set with randomized concentrations. This is to ensure the proper settings for the multivariate regression algorithms. For each dataset, we generated 100 samples in the test set and varied the sample size in the training set from 6 to 24 in steps of 3. During model training, we first performed 3-fold cross validation and parameter grid search within the training set to choose the optimal hyper-parameters for each algorithm (i.e., number of loading vectors in PLS, number of retaining principle components in PCR, and the regularization parameter value in RR respectively). We then refitted the model with the optimal hyper-parameter on the entire training set and applied the resulting model on the test set to calculate the RMSE for the dataset. Since these multivariate regression algorithms have a high variance under the small training sample size regime, we repeated this process a number of times on independently generated training and test sets and report its performance based on aggregated statistics across these independent runs. In Figure 6, we show the average RMSE across 100 independently simulated datasets for the three multivariate regression algorithms as the number of mixture training data varies for different $N_I$ at $\sigma = 3$. The error bars in the plots represent the standard deviation of RMSE across the 100 independent runs. As comparison, we also show the average RMSE for our algorithm across 10 test sets each consisting of 100 mixture spectra in the same plots. The shaded area around the error line indicates the standard deviation of RMSE across the 10 test sets. Since our algorithm does not use the mixture data for training, the error line stays horizontal across the axis for mixture training sample size.

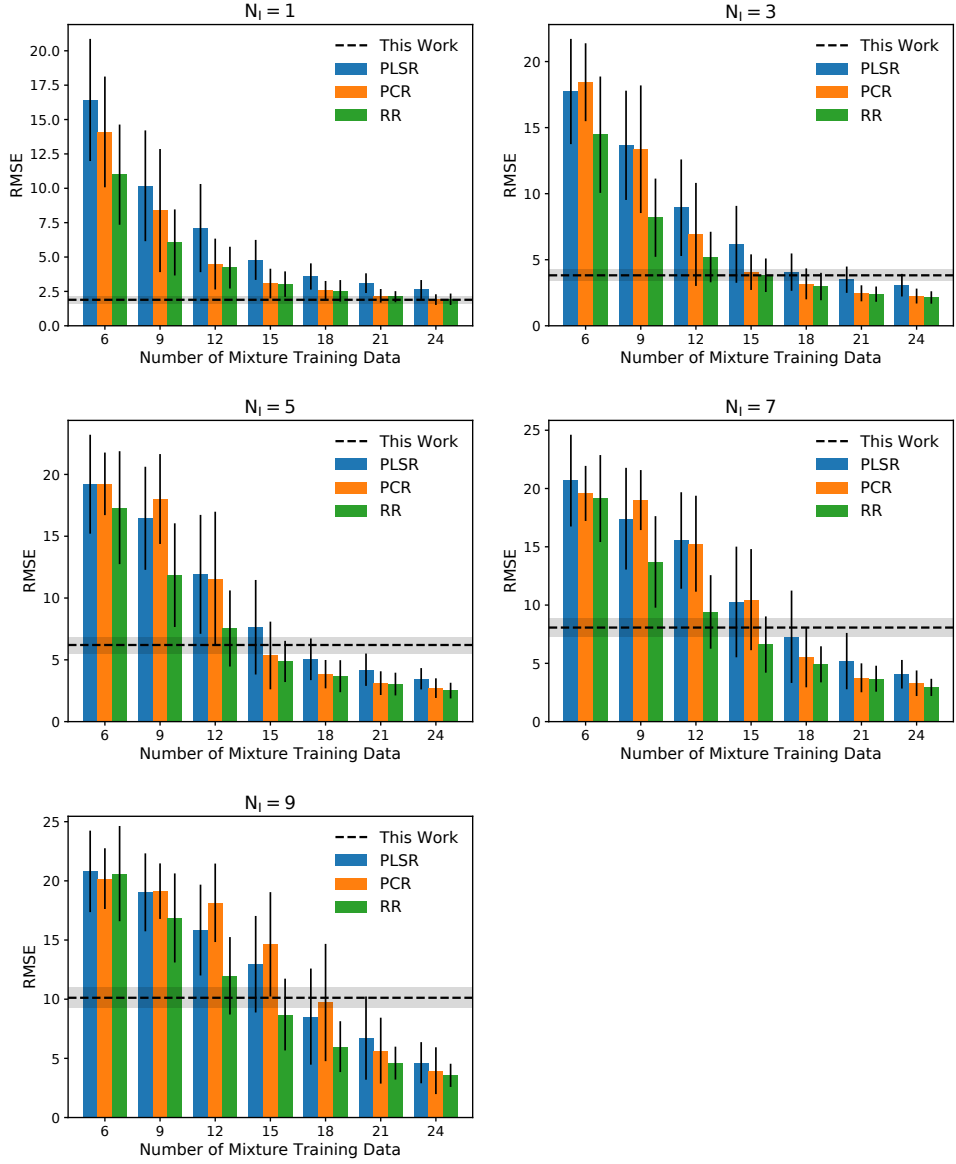

Figure 6: Error plots for our algorithm and three other multivariate regression algorithms with various mixture training data sizes for different $N_I$. The error bars in the bar plots indicate the standard deviation of RMSE for the multivariate regression algorithms across 100 independently simulated datasets each consisting of 100 test spectra. The horizontal shaded area around the dotted line indicates the standard deviation of RMSE for our algorithm across 10 independently simulated datasets each consisting of 100 test spectra. $N_I$ is the number of co-existing interfering analytes in the mixture in our simulations.

| Training Data Size | 6 | 9 | 12 | 15 | 18 | 21 | 24 |
|---|---|---|---|---|---|---|---|
| $N_I = 1$ | | | | | | | |
| This Work | | | | $1.9 \pm 0.3$ | | | |
| PLSR | $16.4 \pm 4.4$ | $10.2 \pm 4.0$ | $7.1 \pm 3.2$ | $4.8 \pm 1.5$ | $3.6 \pm 1.0$ | $3.1 \pm 0.7$ | $2.6 \pm 0.7$ |
| PCR | $14.1 \pm 4.0$ | $8.4 \pm 4.5$ | $4.5 \pm 1.8$ | $3.1 \pm 1.1$ | $2.6 \pm 0.7$ | $2.2 \pm 0.5$ | $1.9 \pm 0.4$ |
| RR | $11.0 \pm 3.6$ | $6.1 \pm 2.4$ | $4.2 \pm 1.5$ | $3.0 \pm 0.9$ | $2.5 \pm 0.8$ | $2.1 \pm 0.4$ | $1.9 \pm 0.4$ |
| $N_I = 3$ | | | | | | | |
| This Work | | | | $3.8 \pm 0.4$ | | | |
| PLSR | $17.7 \pm 4.0$ | $13.7 \pm 4.1$ | $8.9 \pm 3.7$ | $6.2 \pm 2.9$ | $4.1 \pm 1.4$ | $3.5 \pm 1.0$ | $3.1 \pm 0.9$ |
| PCR | $18.4 \pm 2.9$ | $13.4 \pm 4.8$ | $6.9 \pm 3.9$ | $4.1 \pm 1.3$ | $3.2 \pm 1.2$ | $2.5 \pm 0.6$ | $2.3 \pm 0.6$ |
| RR | $14.5 \pm 4.4$ | $8.2 \pm 3.0$ | $5.2 \pm 1.9$ | $3.8 \pm 1.3$ | $3.0 \pm 1.0$ | $2.4 \pm 0.6$ | $2.1 \pm 0.5$ |
| $N_I = 5$ | | | | | | | |
| This Work | | | | $6.2 \pm 0.7$ | | | |
| PLSR | $19.2 \pm 4.0$ | $16.5 \pm 4.2$ | $11.9 \pm 4.8$ | $7.6 \pm 3.8$ | $5.0 \pm 1.7$ | $4.2 \pm 1.3$ | $3.5 \pm 0.9$ |
| PCR | $19.2 \pm 2.5$ | $18.0 \pm 3.6$ | $11.6 \pm 5.4$ | $5.4 \pm 2.7$ | $3.8 \pm 1.1$ | $3.1 \pm 1.0$ | $2.7 \pm 0.8$ |
| RR | $17.3 \pm 4.6$ | $11.9 \pm 4.2$ | $7.5 \pm 3.1$ | $4.9 \pm 1.7$ | $3.7 \pm 1.3$ | $3.0 \pm 0.9$ | $2.5 \pm 0.6$ |
| $N_I = 7$ | | | | | | | |
| This Work | | | | $8.1 \pm 0.8$ | | | |
| PLSR | $20.7 \pm 3.9$ | $17.4 \pm 4.4$ | $15.5 \pm 4.1$ | $10.3 \pm 4.7$ | $7.3 \pm 4.0$ | $5.2 \pm 2.4$ | $4.1 \pm 1.2$ |
| PCR | $19.6 \pm 2.4$ | $19.0 \pm 2.6$ | $15.3 \pm 4.1$ | $10.5 \pm 4.3$ | $5.5 \pm 2.6$ | $3.8 \pm 1.2$ | $3.3 \pm 1.1$ |
| RR | $19.1 \pm 3.7$ | $13.7 \pm 3.9$ | $9.4 \pm 3.2$ | $6.6 \pm 2.4$ | $4.9 \pm 1.5$ | $3.7 \pm 1.1$ | $2.9 \pm 0.7$ |
| $N_I = 9$ | | | | | | | |
| This Work | | | | $10.1 \pm 0.9$ | | | |
| PLSR | $20.8 \pm 3.4$ | $19.0 \pm 3.3$ | $15.8 \pm 3.8$ | $13.0 \pm 4.1$ | $8.5 \pm 4.1$ | $6.7 \pm 3.5$ | $4.6 \pm 1.7$ |
| PCR | $20.2 \pm 2.6$ | $19.1 \pm 2.3$ | $18.1 \pm 3.3$ | $14.6 \pm 4.4$ | $9.7 \pm 5.0$ | $5.7 \pm 2.8$ | $4.0 \pm 2.0$ |
| RR | $20.6 \pm 4.0$ | $16.9 \pm 3.8$ | $12.0 \pm 3.3$ | $8.7 \pm 3.0$ | $6.0 \pm 2.1$ | $4.6 \pm 1.4$ | $3.6 \pm 1.0$ |

Table 2: RMSE values for our algorithm and three other multivariate regression algorithms with various mixture training data sizes for different $N_I$ (also plotted in Figure 6). The number before/after the $\pm$ sign indicates the mean/standard deviation of RMSE across independent runs. $N_I$ is the number of co-existing interfering analytes in the mixture in our simulations.

As shown in these plots, for all the three multivariate regression algorithms, the prediction error for the test set decreases monotonically as a function of the number of mixture training data. This is expected as with more training data, it is more likely for these algorithms to effectively capture the sample subspace of the mixture data during the training process, thereby increasing the regression accuracy. Under small training sample size regime with less than $\approx 15$ training spectra, there is a clear advantage of our algorithm in terms of both estimation accuracy and consistency under all testing situations. On the contrary, once there are enough training data to accurately construct the regression models, our algorithm is unable to match their performance and the accuracy gap widens with more training data. Under low interfering conditions, our algorithm remains competitive across the range of the training sample size change. This is due to the fact that with low interfering conditions, it is more likely to unambiguously resolve analyte peak assignment and accurately estimate peak amplitudes, resulting in near-optimal quantification.

Although a non-negligible accuracy gap is present for the high interfering conditions in Figure 6, we note here that with perfect linearity and fixed mixture components with concentrations generated from uniform distributions in both the training and test sets, our numerical experiment was constructed such that the conventional multivariate regression algorithms were set to achieve near-optimal performance under situations with moderate-to-large training data volume. In practice, apart from experimentation and instrumentation-related issues as described in Wolthuis et al. (2006), the performance of these multivariate regression algorithms are more critically dependent on the quality of the training data, including the size of the training data as well as the measurement conditions for the training and test data. In general it is desirable to select training data that are most representable to the mixture conditions of the test data with sufficient concentration variabilities for critical analytes (Whelan et al., 2012). These requirements can be difficult to satisfy without substantial resources being allocated to the training data collection and verification process. In addition, mixture environment may introduce undesirable effect to the spectral signal. For example, it is known that the Raman peak may shift with environmental pH for many chemicals such as certain amino acids (Mesu et al., 2005). Peak shifts introduce non-linearity into the spectral basis and may reduce estimation accuracy for linear regression algorithms. It is therefore desirable to maintain the pH of the environment for both the training and test data during the measurement for PLSR-like linear regression algorithms (Lee et al., 2004). In contrast, our algorithm is less sensitive to these various requirements so long as the target analyte spectrum stays the same in the mixture comparing to its reference measurement in native form. Therefore, in practice our algorithm may still be comparable or even outperform these multivariate regression algorithms with larger training data volume depending on the nature of the measurement.

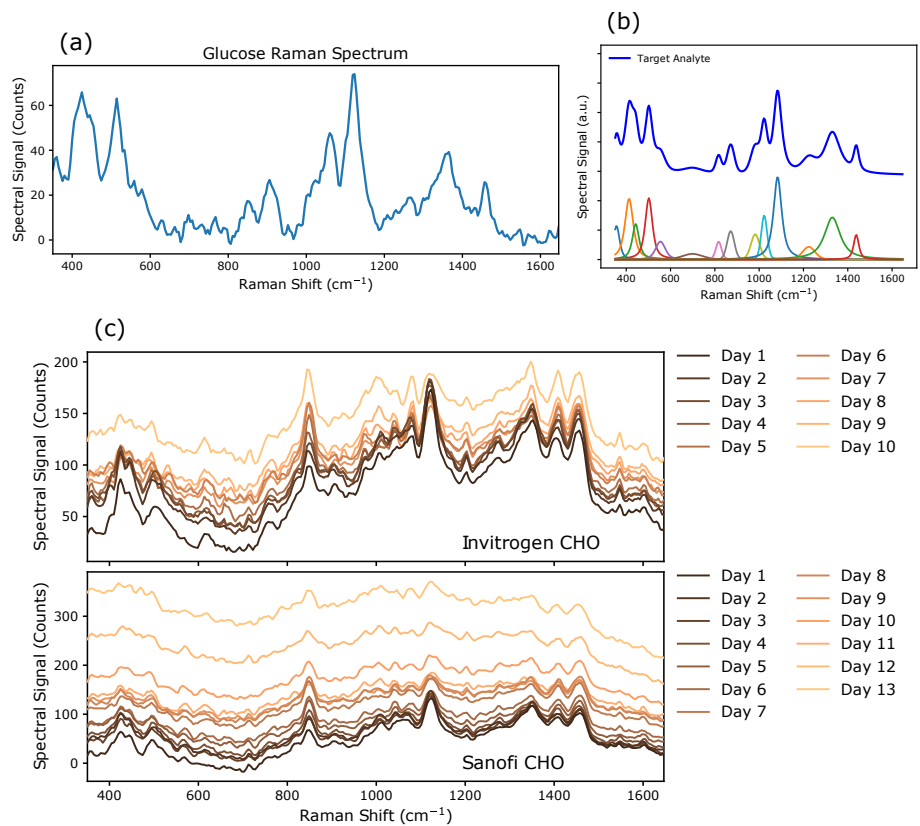

Figure 7: (a) Glucose Raman spectrum measured at 40 mM with our system. (b) Peak decomposition for the glucose Raman spectrum in (a). (c) Raman spectra of the CHO culture supernatant for the Invitrogen (above) and Sanofi (below) CHO cell lines. Preprocessings described in the text had been applied to the raw spectral data and water Raman background was subtracted prior to the plots.

*3.4. Experimental Data Study*

Following the numerical experiments, we further tested our algorithm on experimental Raman spectroscopy data collected for biopharmaceutical applications. Spontaneous Raman spectra were collected to monitor the concentration of the main carbon source, glucose, in the growth environment during the fermentation process of Chinese hamster ovary (CHO) cells, which are the most widely used expression systems for industry production of recombinant protein therapeutics. The initial CHO growth medium included all the nutrients required by the cells such as the necessary carbon sources, nitrogen sources, salts and trace elements. As the fermentation advanced, nutrients were consumed by the cells and metabolites were being produced and released into the growth environment. Therefore, the culture environment represented a complex aqueous mixture and was changing constantly over the fermentation process. Knowledge of key nutrients such as glucose during the fermentation process through an on-line measurement such as Raman spectroscopy can help regulate the fermentation condition for better yield or reproducibility (Whelan et al., 2012). During our experiment, supernatant from the culture material was collected on a daily basis. Raman spectra of the collected supernatant were immediately measured with a confocal Raman spectroscopy system at 830 nm excitation wavelength. Meanwhile, HPLC measurement was used to obtain the reference concentrations for glucose in the supernatant samples. The instrumentation and experimental setup are described in more details in Singh et al. (2015). Two independent experiments with different CHO cell lines from Invitrogen and Sanofi were carried out respectively. In addition to Raman spectra from the supernatant samples, Raman spectrum for pure glucose dissolved in solution was also collected with the same instrument.

The fermentation experiments lasted a total of 10 days for the Invitrogen CHO cell line and 13 days for the Sanofi CHO cell line. Therefore, 10 and 13 supernatant Raman measurements were collected for these two cell lines respectively. The reference Raman spectrum for pure glucose solution was taken at 40 mM concentration. For each set of the Raman measurement, 10 repeated spectra were collected in sequence. As spectral preprocessing, we first took the median across the 10 measurements for each spectral data point for cosmic ray removal and noise reduction. Afterwards, a 21-point Savitzky-Golay filter with a polynomial order of 3 was applied across the spectral dimension to further enhance the spectral signal-to-noise ratio (SNR). A spectral window from 350 $cm^{-1}$ to 1650 $cm^{-1}$, which covered all the major Raman peaks in glucose and CHO Raman spectra, was selected for further processing. Finally, a direct subtraction of Raman spectrum measured with water was carried out to remove background Raman signals from water as well as the optical components along the light path. The processed spectra for glucose as well as the two CHO cell lines are shown in Figure 7. The mixture environments for the two cell lines were different due to the fact that the growth media for these two cell lines had different compositions. This results in the differences in the corresponding Raman spectra shown in Figure 7 (c). The overall baseline drifts over days for each cell line were likely caused by the changing refractive index of the supernatant due to its composition change over the course of the fermentation process.

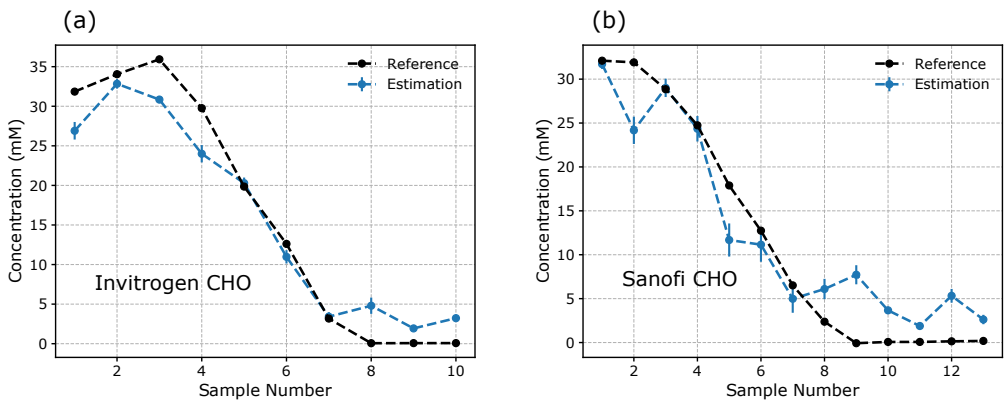

Figure 8: Plots for glucose estimation with Raman spectroscopy for the Invitrogen (left) and Sanofi (right) CHO cell line measurements. The estimated values were obtained as the mean of 10 independent algorithm runs. The error bars indicate the standard error of the mean (SEM) for the estimations across the runs. The reference values were obtained with independent HPLC measurements.

| Cell Line | RMSE (mM) | MAE (mM) | $R^2$ |
|---|---|---|---|
| Invitrogen CHO | 3.5 | 2.9 | 0.94 |
| Sanofi CHO | 4.2 | 3.3 | 0.89 |

Table 3: Estimation results for our algorithm with Raman spectroscopy for the Invitrogen and Sanofi CHO cell line measurements.

We applied our algorithm with the same modeling parameter settings as with the simulated data on the measured glucose and CHO Raman spectra. The average of 10 algorithm runs is plotted in Figure 8 together with the HPLC reference measurements for both cell lines. The error bars in the plots indicate the standard error of the mean (SEM) of the estimation runs. The HPLC measurements were estimated to have $\pm 0.5$ mM accuracy. Overall our algorithm shows a consistent and reliable estimation of glucose, as shown in Table 3, with RMSE of 3.5 mM, mean absolute error (MAE) of 2.9 mM, and $R^2$ of 0.94 for the Invitrogen CHO measurement, and RMSE of 4.2 mM, MAE of 3.3 mM, and $R^2$ of 0.89 for the Sanofi CHO measurement. The error is comparable with the 3-$\sigma$ limit of detection for pure glucose solution with our measurement system, which was $\approx 6$ mM based on peak-SNR estimation. Comparing with conventional PLSR-like multivariate regression algorithms, our algorithm only requires additional measurements of pure glucose solution and water. Otherwise additional experimental runs need to be planned in order to accumulate enough data for model training and validation. With resource-intensive applications like industrial fermentation, a sample-efficient approach like our algorithm can significantly reduce the research cost and development cycle. It is worth noting that it is also possible to explicitly measure spectra for all constituent analytes in the mixture and use ordinary least squares (OLS) regression to quantify analyte concentrations without acquiring additional mixture training data (Lee et al., 2004; Singh et al., 2015). However, the library spectra collection process can be labor-intensive. In addition, it is usually difficult to know all the mixture constituents ahead of time in a general setting. Therefore, in practice, our algorithm has advantages in terms of both performance competitiveness as well as looser requirement on training or additional measurements.

## 4. Discussions and Conclusions

In summary, we have developed a two-stage quantification algorithm with the Bayesian modeling framework and the RJMCMC computation. We tested our algorithm on both simulated as well as experimentally collected spontaneous Raman spectroscopy datasets to validate its usage. The successful quantification of glucose concentration in a complex aqueous cell culture environment without any mixture training data suggests its promising potential for applications involving Raman spectral analysis.

In practice, collecting high quality Raman spectroscopy training datasets with reference measurements in sufficient volume for multivariate regression algorithms can be a long, challenging, and labor/resource-intensive process for many application disciplines. In addition, due to the intertwined nature of statistical data analysis and experimental design in chemometrics, timely quantitative feedback can often impact aspects of experimental design in a significant way. An analyte quantification algorithm without any requirement on mixture training data such as the one developed in this work is therefore desirable in many scenarios. As a result, we envision our algorithm to be a complementary tool to the multivariate regression algorithms for quantification analysis with Raman spectroscopy datasets.

Although RJMCMC computation was chosen as a building block for this work, other statistical sampling methods or approximate Bayesian inference techniques with a suitable model selection criterion can be used within the general two-stage processing framework as well. These methods can potentially have advantages in terms of computational speed and therefore may be suitable for scalable applications. We leave this aspect as a potential future direction for our work.

## Acknowledgements

We'd like to thank Dr. Gajendra P. Singh for providing the CHO culture Raman spectroscopy dataset, which was acquired under support from Sanofi S.A. We'd also like to thank Dr. William F. Herrington for helpful discussions about Raman spectroscopy acquisition and processing.

## Appendix A. Gibbs Updates

With Gibbs sampling, $g$ can be updated with an inverse-gamma distribution as

$$g|\cdots \sim \mathcal{IG}\left(a_g + \frac{k}{2}, b_g + \frac{1}{2\sigma^2}(\boldsymbol{\beta}_k - \boldsymbol{\beta}_{k,0})^\intercal \mathbf{X}_k^\intercal(\boldsymbol{\theta}_P)\mathbf{X}_k(\boldsymbol{\theta}_P)(\boldsymbol{\beta}_k - \boldsymbol{\beta}_{k,0})\right).$$

$\Lambda$ can be updated as

$$\Lambda|\cdots \sim \mathcal{G}a(a_\Lambda + k_P, b_\Lambda + 1).$$

Denoting $\hat{\boldsymbol{\beta}}_k = \left(\mathbf{X}_k^\intercal(\boldsymbol{\theta}_P)\mathbf{X}_k(\boldsymbol{\theta}_P)\right)^{-1}\mathbf{X}_k^\intercal(\boldsymbol{\theta}_P)\mathbf{y}$ as the maximum likelihood (ML) estimation of $\boldsymbol{\beta}_k$ and $s^2 = (\mathbf{y} - \mathbf{X}_k(\boldsymbol{\theta}_P)\hat{\boldsymbol{\beta}}_k)^\intercal(\mathbf{y} - \mathbf{X}_k(\boldsymbol{\theta}_P)\hat{\boldsymbol{\beta}}_k)$ as the squared residue of the ML estimation, we define

$$\tilde{b}_{\sigma^2} = \frac{s^2}{2} + \frac{1}{2(g+1)}(\hat{\boldsymbol{\beta}}_k - \boldsymbol{\beta}_{k,0})^\intercal \mathbf{X}_k^\intercal(\boldsymbol{\theta}_P)\mathbf{X}_k(\boldsymbol{\theta}_P)(\hat{\boldsymbol{\beta}}_k - \boldsymbol{\beta}_{k,0}). \qquad (A.1)$$

With this definition, our posterior for $\sigma^2$ and $\boldsymbol{\beta}_k$ can be updated as

$$\sigma^2|\cdots \sim \mathcal{IG}\left(\frac{N}{2}, \tilde{b}_{\sigma^2}\right),$$

and

$$\boldsymbol{\beta}_k|\sigma^2,\cdots \sim \mathcal{N}\left(\frac{1}{g+1}\left(\boldsymbol{\beta}_{k,0} + g\hat{\boldsymbol{\beta}}_k\right), \frac{g}{g+1}\sigma^2\left(\mathbf{X}_k^\intercal(\boldsymbol{\theta}_P)\mathbf{X}_k(\boldsymbol{\theta}_P)\right)^{-1}\right). \qquad (A.2)$$

# Appendix B. Calculation of the RJMCMC Move Acceptance Probabilities

For within-dimensional moves where the dimensionality of the variable space stays the same and $k'_P = k_P$, we can use symmetric random walks in $\boldsymbol{\theta}_P$ to generate $\boldsymbol{\theta}'_P$. This is essentially the Metropolis algorithm and $A(\mathbf{x}, \mathbf{x}')$ is the ratio between the posterior density function for the proposed state $\mathbf{x}'$ and the current state $\mathbf{x}$ as

$$A_{\text{within}}(\mathbf{x}, \mathbf{x}') = \left(\frac{\tilde{b}'_{\sigma^2}}{\tilde{b}_{\sigma^2}}\right)^{-\frac{N}{2}} \prod_{i=1}^{k_P} \mathbb{1}_{[l_{\min}, l_{\max}]}(l'_i)$$

$$\left[\prod_{i=1}^{k_P} \left(\frac{w'_i}{w_i}\right)^{-a_w - 1}\right] \exp\left\{-b_w \sum_{i=1}^{k_P} \left(\frac{1}{w'_i} - \frac{1}{w_i}\right)\right\} \prod_{i=1}^{k_P} \mathbb{1}_{[\rho_{\min}, \rho_{\max}]}(\rho'_i).$$

For trans-dimensional sampling, four individual moves have been used in this study, namely the birth, death, split, and merge moves. For the birth move with $k'_P = k_P + 1$, a peak is generated with $\theta_b = (l_b, w_b, \rho_b)$, where $l_b$ is randomly drawn from $[l_{\min}, l_{\max}]$, $w_b$ is randomly drawn from density function $p_{w_b}(w_b)$, and $\rho_b$ is randomly drawn from $[\rho_{\min}, \rho_{\max}]$. For $p_{w_b}(w_b)$, we choose to use the empirically fitted inverse-gamma distribution that is described in Section 3.1. With this, $A_{\text{birth}}(\mathbf{x}, \mathbf{x}')$ can be shown as

$$A_{\text{birth}}(\mathbf{x}, \mathbf{x}') = \left(\frac{\tilde{b}'_{\sigma^2}}{\tilde{b}_{\sigma^2}}\right)^{-\frac{N}{2}} \frac{\Lambda}{k'_P}(g+1)^{-\frac{1}{2}} \frac{b_w^{a_w}}{\Gamma(a_w)} w_b^{-a_w - 1} e^{-\frac{b_w}{w_b}} \frac{1}{k'_P} \frac{1}{p_{w_b}(w_b)}.$$

Meanwhile, for the reversed process of randomly killing an existing peak with peak variables $\theta_d = (l_d, w_d, \rho_d)$, we have $k'_P = k_P - 1$ and

$$A_{\text{death}}(\mathbf{x}, \mathbf{x}') = \left(\frac{\tilde{b}'_{\sigma^2}}{\tilde{b}_{\sigma^2}}\right)^{-\frac{N}{2}} \frac{k_P}{\Lambda}(g+1)^{\frac{1}{2}} \frac{\Gamma(a_w)}{b_w^{a_w}} w_d^{a_w + 1} e^{\frac{b_w}{w_d}} k_P \, p_{w_b}(w_d).$$

For the split move, a random peak is selected and split into two adjacent peaks. Assume that the selected peak has the peak variables as $\theta_s = (l_s, w_s, \rho_s)$, we split the peak into two peaks with $\theta_s^+ = (l_s^+, w_s^+, \rho_s^+)$ and $\theta_s^- = (l_s^-, w_s^-, \rho_s^-)$ as

$$l_s^+ = l_s + \delta_l u_l, \quad l_s^- = l_s - \delta_l u_l,$$

$$w_s^+ = w_s + \delta_w u_w, \quad w_s^- = w_s - \delta_w u_w,$$
$$\rho_s^+, \rho_s^- \sim \mathcal{U}(0, 1),$$

where $\delta_l$ and $\delta_w$ are the hyper-parameters to specify the variable split ranges, $u_l \sim \mathcal{U}(0, 1)$, and $u_w \sim \mathcal{U}(-1, 1)$. The corresponding $A_{\text{split}}(\mathbf{x}, \mathbf{x}')$ with $k_P' = k_P + 1$ is

$$A_{\text{split}}(\mathbf{x}, \mathbf{x}') = \left(\frac{\tilde{b}'_{\sigma^2}}{\tilde{b}_{\sigma^2}}\right)^{-\frac{N}{2}} \frac{\Lambda}{k_P'}(g+1)^{-\frac{1}{2}} \frac{1}{\Delta_l} \frac{b_w^{a_w}}{\Gamma(a_w)} \left(\frac{w_s^+ w_s^-}{w_s}\right)^{-a_w-1} e^{-b_w\left(\frac{1}{w_s^+} + \frac{1}{w_s^-} - \frac{1}{w_s}\right)} 8\delta_l \delta_w.$$

For the reversed merge move, the peak variables $\theta_m^+ = (l_m^+, w_m^+, \rho_m^+)$ and $\theta_m^- = (l_m^-, w_m^-, \rho_m^-)$ from the two selected adjacent peaks are merged into $\theta_m = (l_m, w_m, \rho_m)$ as

$$l_m = \frac{1}{2}\left(l_m^+ + l_m^-\right), \quad w_m = \frac{1}{2}\left(w_m^+ + w_m^-\right), \quad \rho_m \sim \mathcal{U}(0, 1).$$

With $k_P' = k_P - 1$, $A_{\text{merge}}(\mathbf{x}, \mathbf{x}')$ can be calculated as

$$A_{\text{merge}}(\mathbf{x}, \mathbf{x}') = \left(\frac{\tilde{b}'_{\sigma^2}}{\tilde{b}_{\sigma^2}}\right)^{-\frac{N}{2}} \frac{k_P}{\Lambda}(g+1)^{\frac{1}{2}} \Delta_l \frac{\Gamma(a_w)}{b_w^{a_w}} \left(\frac{w_m^+ w_m^-}{w_m}\right)^{a_w+1} e^{b_w\left(\frac{1}{w_m^+} + \frac{1}{w_m^-} - \frac{1}{w_m}\right)} \frac{1}{8\delta_l \delta_w}.$$

For the trans-dimensional move pairs, we make sure that the moves within each pair are reversible. For the split and merge move pair, this means that if a split move creates two peaks that are not adjacent in the current spectrum, or if the selected adjacent peaks in the merge move have larger variable differences in $l$ and $w$ than those that are allowed in the split move, we would discard the proposal to ensure reversibility.